\documentclass{PoS}
\usepackage{multirow}
\def\lsim{\mathrel{\rlap{\lower4pt\hbox{\hskip1pt$\sim$}}
    \raise1pt\hbox{$<$}}}                
\def\gsim{\mathrel{\rlap{\lower4pt\hbox{\hskip1pt$\sim$}}
    \raise1pt\hbox{$>$}}}                

\title{Physics and Astrophysics Opportunities with Supernova Neutrinos}

\ShortTitle{Supernova Neutrinos: A Review}

\author{\speaker{Basudeb Dasgupta}\\
        CCAPP, Ohio State University, 191 W. Woodruff Avenue, Columbus 43210 U.S.A\\
        E-mail: \email{dasgupta.10@osu.edu}}

\abstract{Neutrinos emitted from a supernova encode useful information about neutrino physics and astrophysics. Interpreting the neutrino signal depends crucially on understanding neutrino production, flavor mixing during propagation, and detection. In this talk, we review the physics potential of a SN neutrino observation.}

\FullConference{35th International Conference of High Energy Physics - ICHEP2010,\\
		July 22-28, 2010\\
		Paris France}

\begin{document}

\section{Introduction}
Neutrinos from a core-collapse supernova (SN)~\footnote{Neutrinos refer to both neutrinos and antineutrinos. Only neutrinos will be written as $\nu$. Similarly, antineutrinos will be written as $\bar{\nu}$. Also, SN refers to core-collapse supernovae only.} provide a rare and valuable physics opportunity~\cite{Dighe:2008dq}. The gargantuan fluxes and low interaction rates of SN neutrinos, makes them ideal candidates to probe neutrino mixing properties and study the extreme conditions in the depths of a star.  Observation of the SN $1987$A in neutrinos opened the field of neutrino astronomy~\cite{Koshiba:1992yb}, and confirmed our overall understanding of SN neutrino physics~\cite{Raffelt:1996wa}. With present and planned detectors~\cite{Scholberg:2010zz}, a galactic SN is expected to result in a high-statistics detection. This would lead to  significant advances in neutrino physics and SN astrophysics, if we can disentangle the relevant information.

Decoding the neutrino signal requires a detailed understanding of neutrino production, mixing during propagation, and detection. For neutrinos, the mixing scenario is reasonably well-determined, except the value of $\theta_{13}$, the sign of $\Delta m^2_{\rm atm}$, and the CP-violating phase~\cite{GonzalezGarcia:2010er}, and the main detection channels are well-calibrated. However for SNe, initial neutrino fluxes and spectra predicted from SN theory have a significant variance~\cite{Simulations}. Similarly, the stellar conditions are poorly constrained. The signal interpretation is plagued by these uncertainties, and the strategy for disentangling information from a SN neutrino signal must rely on generic features that are insensitive to model assumptions.

Some information is obtained directly, e.g. the direction and time structure of the event, and the above mentioned uncertainties do not affect our inferences.  But a lot more information, e.g. flavor dependent energy spectra of the neutrinos, clues to the unknown neutrino parameters, and some signatures of stellar dynamics, is encoded in a flavor dependent way. A detailed treatment of neutrino mixing is required to extract that information. Traditional analyses of neutrino mixing for SN took into account interactions with matter, through the Mikheev-Smirnov-Wolfenstein  (MSW) effect. This MSW-based paradigm was believed to be complete, and numerous results followed. See the review~\cite{Dighe:2004xy} for the traditional expectations. However, that picture was incomplete. Deep in the SN, neutrino densities are large enough to make their collective interactions extremely important. The mixing angles are highly matter suppressed, and one may expect no flavor conversion in that region. However, this naive expectation is incorrect. Neutrino-neutrino interactions entangle the flavor evolution of all neutrinos and create an instability in the flavor composition. Thus flavor conversions take place even for extremely small mixing angles, with a rich phenomenology.  See the recent review~\cite{Duan:2010bg} on these so-called  ``Collective effects''. In this talk, we summarize the main aspects of SN neutrinos, focussing on the collective effects and their impact on the physics potential of SN neutrino observations.

\section{Neutrino production and primary fluxes}
Neutrinos are produced in the core of the star, and remain trapped inside their respective energy and flavor dependent ``neutrinospheres'' at densities of $\sim10^{10}$g/cc. As the star gravitationally collapses, the core reaches nuclear density and becomes incompressible. A hydrodynamic shock travels outwards, and as it passes through the neutrinospheres, a $\nu_e$ ``neutronization burst'', lasting $\sim10$ ms, is emitted due to rapid electron capture on dissociated nuclei. Material continues to fall onto the star for the following $\sim100$ ms in the ``accretion'' phase, gets heated and emits neutrinos of all species. The object below the shock wave, the protoneutron star, then cools down with the emission of neutrinos, over a time period of approximately $10$ s in the Kelvin-Helmholtz ``cooling'' phase. The eventual explosion of the star involves damping of the original shock wave, its rejuvenation by neutrino heating, and a ``delayed'' explosion. See the review~\cite{Janka:2006fh} for a more detailed account.

Almost all of the gravitational binding energy of the star $\sim10^{53}$ erg is converted to neutrinos. The proto-neutron star acts almost like a thermal neutrino source
with flavor dependent fluxes. The ``primary fluxes'' $F^0_{\nu_\alpha}$ are parametrized by total number fluxes $\Phi_{\nu_\alpha}$, average energies 
$\langle E_{\nu_\alpha} \rangle$, and spectral parameters that characterize deviation from thermal spectra. Values of the parameters are model dependent~\cite{Keil:2003sw}. However, $\nu_e$ are expected to have $\langle E_{\nu_e} \rangle\approx 10-12$ MeV, while $\bar{\nu}_e$ are expected to have slightly higher average energy $\langle E_{\bar{\nu}_e} \rangle\approx 12-15$ MeV, owing to its fewer interactions with the neutron rich matter. Interactions are almost identical for $\nu_\mu$, $\nu_\tau$, and their antiparticles, so it is convenient to work in terms of the three ``flavors'': $\nu_e, \nu_x\equiv\cos\theta_{23}\nu_{\mu}-\sin\theta_{23}\nu_{\tau}$ and $\nu_y\equiv\cos\theta_{23}\nu_{\mu}+\sin\theta_{23}\nu_{\tau}$. Clearly, $\nu_x$ and $\nu_y$ have identical primary fluxes $F^0_{\nu_{x}}$ and average energies $\langle E_{\nu_{x}} \rangle\approx 15-25$ MeV. The luminosities $L_{\nu_{\alpha}}$ are quasi-equipartitioned, with $L_{\nu_\alpha}\sim10^{52}$ erg/s, depending somewhat on the phase of the explosion.

\section{Neutrino flavor conversion}
The nature of neutrino flavor conversions depends on an interplay of neutrino oscillation frequency $\omega=\Delta m^2/(2E)$ with the matter potential $\lambda=\sqrt{2} G_F n_e$ due to background electrons, and with the collective neutrino potential $\mu=\sqrt{2} G_F(1-\cos{\theta}) n_{\nu +{\bar{\nu}}}$ generated by other neutrinos. Thus enhanced conversion can happen either due to matter effects, or due to the neutrino potential, or an interplay of the two. In typical supernovae, the matter potential falls as $n_{e} \propto 1/r^3$ with radius, whereas the collective potential falls off faster as $n_{\nu+\bar{\nu}}\langle 1-\cos\theta\rangle\propto1/r^4$. So, when the neutrinos travel outward from the SN core, they first experience collective effects, and then matter effects, which may  be modified by shock wave effects. After they leave the SN, the mass eigenstates travel independently and are detected on Earth as an incoherent superposition. There can be distinctive effects due to additional conversions during propagation inside the Earth.

The final outcome for the $\nu_{e}$ and $\bar{\nu}_e$ fluxes can be written down simply in terms of their overall survival probability $p_{\nu_{e}}$ and $p_{\bar{\nu}_e}$ respectively, i.e. $F_{\nu_e}=p_{\nu_{e}}F^0_{\nu_e}+(1-p_{\nu_{e}})F^0_{\nu_x}$, and similarly for $\bar{\nu}_e$. The values of $p_{\nu_{e}}$ and $p_{\bar{\nu}_e}$ are given in Table~\ref{Table1}. In the remainder of this section, we discuss the main aspects of these flavor conversions.
 
 \subsection{Collective oscillations due to neutrino-neutrino interactions}
The neutrino density creates a potential that is not flavor diagonal \cite{Pantaleone:1992eq}; $n_{\nu},n_{\bar{\nu}}$ are density matrices in flavor space and depend on the flavor composition of the entire neutrino ensemble! Flavor evolution of such dense relativistic neutrino gases~\cite{Sigl:1992fn} can be understood to good accuracy without considering many-particle effects~\cite{Friedland:2003dv}. Recent simulations in spherical symmetry showed that the collective oscillations affect neutrino flavor conversions substantially~\cite{Duan:2005cp}~\cite{Duan:2006an}. The main features observed were large flavor conversions for inverted hierarchy, and a surprisingly mild dependence on the mixing angle and the matter density. 

These features can be understood analytically. A dense gas of neutrinos displays collective flavor conversion~\cite{Kostelecky:1994dt}, i.e. neutrinos of all energies oscillate almost in phase, through synchronized~\cite{Pastor:2001iu}/parametrically resonant~\cite{Raffelt:2008hr}/bipolar oscillations~\cite{Hannestad:2006nj}~\cite{Duan:2007mv}. The effect of the bipolar oscillations with a decreasing $\mu$ is a partial or complete swapping of the energy spectra of two neutrino flavors~\cite{Raffelt:2007cb}~\cite{Dasgupta:2009mg}.  The ``$1-\cos\theta$'' structure  of weak interactions can give rise to a dependence of flavor evolution on the neutrino emission angle~\cite{Duan:2006an} or even decoherence~\cite{Raffelt:2007yz}. For a realistic asymmetry between $\nu$ and $\bar{\nu}$ fluxes, such angle-dependent effects are likely to be small~\cite{EstebanPretel:2007ec}~\cite{Fogli:2007bk}.  Even non-spherical source geometries can often be captured by an effective single-angle approximation~\cite{Dasgupta:2008cu} in the coherent regime. Three-flavor effects can be factorized into oscillations driven by the atmospheric parameters on one hand and solar parameters on another~\cite{Dasgupta:2007ws}. Effects of CP violation are suppressed when $\mu$ and $\tau$ are almost equivalent~\cite{Gava:2008rp}. However, any departure from $\mu-\tau$ equivalence triggers collective effects even for a vanishing mixing angle~\cite{Blennow:2008er}~\cite{Dasgupta:2010ae}.

Though the inherent nonlinearity and the presence of multi-angle effects make the analysis rather complicated, the final outcome for the neutrino fluxes turns out to be rather straightforward, at least in the spherically symmetric scenario. Synchronized oscillations with a frequency $\langle \omega \rangle$ take place just outside the neutrinosphere at $r \sim 10-40$ km. These cause no significant flavor conversions since the mixing angle is highly suppressed by the large matter density. A known exception occurs for the burst phase of low-mass supernovae, when the matter density is low~\cite{Duan:2007sh}. In such a situation, neutrinos of all energies undergo MSW resonances together with the same adiabaticity, before collective effects become negligible~\cite{Duan:2008za}~\cite{Dasgupta:2008cd}. At larger radii $r\sim40-100$ km, bipolar or pendular oscillations ${\nu_e}\leftrightarrow {\nu_{x,y}}$ with a higher frequency $\sqrt{2\omega\mu}$ follow. These oscillations are instability driven and thus depend logarithmically~\cite{Hannestad:2006nj} on the the mixing angle or initial misalignment,  occurring where the fluxes for the two flavors are very similar~\cite{Dasgupta:2009mg}. As $\mu$ decreases so that $\langle\omega\rangle\sim\mu$, neutrinos near this instability may relax to the lower energy state. As a result, one finds one or more spectral swaps demarcated by sharp discontinuities or ``spectral splits'' in the oscillated flux.

The final outcome depends on the ordering of initial fluxes. The situation is relatively simple if $L_{\nu_e}\approx L_{\bar{\nu}_e}\gsim L_{\nu_x}$, as is usually expected in the accretion phase. Spectral swaps happen in a simple way, as summarized in~\cite{Dasgupta:2007ws}.  For inverted hierarchy $(\Delta m^2_{\rm atm}<0)$, a swap ${\nu_e}\leftrightarrow {\nu_y}$ above a critical energy $E_{low}\approx10$ MeV is seen. In the normal hierarchy $(\Delta m^2_{\rm atm}>0)$, collective effects do not cause any swapping of neutrino spectra. On the other hand, if $L_{\nu_x}/L_{\nu_e}>1$, as is often predicted for the cooling phase, one finds multiple spectral swaps~~\cite{Dasgupta:2009mg}. For inverted hierarchy, a swap ${\nu_e}\leftrightarrow {\nu_y}$ takes place at intermediate energies ($E_{low}\lsim E \lsim E_{high}$), where $E_{high}\approx 25$~MeV. Furthermore, collective oscillations driven by solar parameters produce a swap ${\nu_e}\leftrightarrow {\nu_x}$ in the high energy spectra~\cite{Friedland:2010sc}~\cite{Dasgupta:2010cd}, unless the fluxes are say, within $10-30\%$ of equipartition~\cite{Choubey:2010up}. For normal hierarchy, a ${\nu_e}\leftrightarrow {\nu_y}$ swap happens at $E\gsim E_{high}$. The antineutrino fluxes also swap in a similar way as the neutrino fluxes, except that the split energies in are lower by $\sim 5$ MeV. The ${\bar{\nu}_e}\leftrightarrow {\bar{\nu}_y}$ swap at intermediate energies fails~\cite{Choubey:2010up} if $\langle E_{\nu_x}\rangle\lsim 18$ MeV, because the adiabaticity is often lower, leading to incomplete swaps~\cite{Friedland:2010sc}. In general, an interplay of various effects produces a rich phenomenology of flavor conversions that remains to be fully understood.

\subsection{Enhanced conversions at MSW resonances and shock waves}
After the collective oscillations, the rest of the flavor evolution is essentially unchanged from the MSW-based paradigm~\cite{Dighe:1999bi}. In the outer layers of the star at $r\sim 1000$ km from the center, the neutrinos encounter the MSW resonances. A resonance at matter densities $10^3-10^5$ g/cc, which correspond to $\Delta m^2_{\rm atm}$, is called an H resonance. When the density is $30-300$ g/corresponding to $\Delta m^2_{\rm sol}$, it is called an L resonance. The H resonance takes place in $\nu$ for the normal hierarchy, and in $\bar{\nu}$ for the inverted hierarchy. The L resonance is always in neutrinos. At the resonance, the neutrinos have a tendency to swap their flavor. The conversion efficiency depends on the gradient of $n_e$ at the MSW resonance, which if large can cause further non-adiabatic flavor conversion. In the static limit of the matter density profile, the H resonance is adiabatic for a large mixing angle $(\sin^2\theta_{13}>10^{-3})$ and non-adiabatic for small mixing angle $(\sin^2\theta_{13}<10^{-5})$. The L resonance is always adiabatic.

When the shock wave passes through the resonance region, it makes the previously adiabatic resonances temporarily non-adiabatic thus changing the survival probability~\cite{Schirato:2002tg}. These shock wave effects on observable neutrino fluxes leave model independent signatures in the energy-time spectra~\cite{Takahashi:2002yj}~\cite{Lunardini:2003eh}~\cite{Fogli:2003dw}~\cite{Tomas:2004gr}. Multiple shock fronts give rise to multiple resonances and result in possible interference effects~\cite{Dasgupta:2005wn}. Stochasticity~\cite{Fogli:2006xy} or turbulence~\cite{Friedland:2006ta} behind the shock wave may depolarize the neutrino ensemble,  and partially wash out shock wave effects. Whenever the survival probability $p_{\nu_e}$ or $p_{\bar{\nu}_e}$ changes between the large $\theta_{13}$ and small $\theta_{13}$ scenarios for a given $\Delta m^2_{\rm atm}$, one finds shock induced non-adiabaticity and the large mixing case behaves temporarily like the small mixing case (See Table~\ref{Table1}).

\subsection{Flavor regeneration in Earth matter}
As the neutrinos leave the star, they travel as independent mass eigenstates and are detected at Earth. Earth matter effects modify on the neutrino fluxes as they pass through the Earth before being detected~\cite{Cribier:1986ak}. Usually, we detect the electron flavor flux. In the presence of Earth effects, whenever the survival probability $p_{\nu_e}$ or $p_{\bar{\nu}_e}$ are not zero or one, they depend on the solar mixing angle $\theta_{12}$ (See Table~\ref{Table1}). For a path-length $L$ inside the Earth this mixing angle becomes an oscillatory function of $L/E$ and one  finds wiggles in the energy spectra for a shadowed detector~\cite{Dighe:2001rs}~\cite{Lunardini:2001pb}~\cite{Takahashi:2001dc}.
\begin{table}[!t]
\caption{Survival probability of $\nu_e$ and $\bar{\nu}_e$ in different phases of a SN explosion and neutrino mass/mixing scenarios. The caveats $[a]$ or $[b]$ refer to cases of low $\langle E_{\nu_x} \rangle$ ($<18$ MeV) or only weakly broken equipartition ($L_{\nu_x}/L_{\nu_e}\approx 1.0-1.3$) respectively, in which case $p_{\nu_e}$ and $p_{\bar{\nu}_e}$ are the same as that at $E\lsim E_{low}$; the $\bar{\nu}_e \leftrightarrow \bar{\nu}_y$ and $e \leftrightarrow x$ swaps fail to take place in an efficient way for those cases respectively. See Sec.3.1 for more details. 
}
\begin{center}
\begin{tabular}{|c|c|c|cc|cc|}
\hline
&&Burst&\multicolumn{2}{c|}{Accretion}&\multicolumn{2}{c|}{Cooling}\\
&&&\multicolumn{2}{c|}{$(L_{\nu_x} \lsim L_{\nu_e}$)}&\multicolumn{2}{c|}{($L_{\nu_x}\gsim L_{\nu_e}$)}\\
\hline
Mass and Mixing&Energy&$p_{\nu_e}$&$p_{\nu_e}$&$p_{\bar{\nu}_e}$&$p_{\nu_e}$&$p_{\bar{\nu}_e}$\\
\hline
\hline
{$\Delta m^2_{\rm atm}>0$ with}&$E\lsim E_{high}$& \multirow{2}{*}{$0$} &\multirow{2}{*}{$0$}&\multirow{2}{*}{$\cos^2\theta_{12}$}&$0$&$\cos^2\theta_{12}$\\
$\sin^2\theta_{13}>10^{-3}$&$E\gsim E_{high}$&&&&$\sin^2\theta_{12}$&$0$\\
\hline
{$\Delta m^2_{\rm atm}>0$ with}&$E\lsim E_{high}$& \multirow{2}{*}{$\sin^2\theta_{12}$} &\multirow{2}{*}{$\sin^2\theta_{12}$}&\multirow{2}{*}{$\cos^2\theta_{12}$}&$\sin^2\theta_{12}$&$\cos^2\theta_{12}$\\
$\sin^2\theta_{13}<10^{-5}$&$E\gsim E_{high}$&&&&$0$&$0$\\
\hline
\multirow{2}{*}{$\Delta m^2_{\rm atm}<0$ with}&$E\lsim E_{low}$&\multirow{3}{*}{$\sin^2\theta_{12}$}&{$\sin^2\theta_{12}$}&{$0$}&$\sin^2\theta_{12}$&$0$\\
\multirow{2}{*}{$\sin^2\theta_{13}>10^{-3}$}&$E_{low} \lsim E \lsim E_{high}$&&\multirow{2}{*}{$0$}&\multirow{2}{*}{$\cos^2\theta_{12}$}&$0$&$\cos^2\theta_{12}~[a]$\\
&$E\gsim E_{high}$&&&&$\cos^2\theta_{12}~[b]$&$\sin^2\theta_{12}~[b]$\\
\hline
\multirow{2}{*}{$\Delta m^2_{\rm atm}<0$ with}&$E\lsim E_{low}$&\multirow{3}{*}{$\sin^2\theta_{12}$}&{$\sin^2\theta_{12}$}&{$\cos^2\theta_{12}$}&$\sin^2\theta_{12}$&$\cos^2\theta_{12}$\\
\multirow{2}{*}{$\sin^2\theta_{13}<10^{-5}$}&$E_{low}\lsim E \lsim E_{high}$&&\multirow{2}{*}{$0$}&\multirow{2}{*}{$0$}&$0$&$0~[a]$\\
&$E\gsim E_{high}$&&&&$\cos^2\theta_{12}~[b]$&$\sin^2\theta_{12}~[b]$\\
\hline
\end{tabular}
\end{center}
{\footnotesize{Note: When survival probability is not zero, one we can get Earth effects. Also, if the survival probabilities differ for large and small $\theta_{13}$, shock effects are seen for corresponding large mixing scenario.}}
\label{Table1}
\end{table}

\section{SN neutrino detection and interpretation}
 
 The rate of SN explosions in our Galaxy is estimated to about $1-3$ per century using a variety of methods~\cite{Diehl:2006cf}. The mean distance is $\sim 10$ kpc, with a fairly large variance of $\sim5$ kpc~\cite{Mirizzi:2006xx}. Thus existing and planned detectors can expect to observe $\sim 10^2-10^6$ neutrinos for a Galactic event. With larger detectors, neutrinos from beyond the immediate neighborhood in the galaxy may also be detected, albeit with low statistics~\cite{Kistler:2008us}. Finally, neutrinos from all the supernovae that have exploded in our past form a diffuse background. See the recent review for an overview~\cite{Beacom:2010kk}.

\subsection{Pointing, timing, and distance}

Detectors like SK can expect to point  within a few degrees of an impending SN,  in advance of the actual explosion~\cite{Beacom:1998fj}~\cite{Tomas:2003xn}. It would be possible to determine the bounce time of the SN within $\sim1$ ms using present detectors~\cite{Pagliaroli:2009qy}~\cite{Halzen:2009sm}. Although a core collapse SN is not a standard candle, the neutronization burst comes close~\cite{Takahashi:2003rn}, and can be used to determine the distance~\cite{Kachelriess:2004ds} to within $5-10\%$. All of this information becomes increasingly useful for a dust-obscured SN. In addition, because the neutrino burst precedes the actual optical display by a few hours to a day,  it can used to provide an early warning signal to astronomers~\cite{Antonioli:2004zb}.

\subsection{Neutrino masses and $\theta_{13}$}
Neutrino physics stands to gain immensely from a SN observation. The neutronization phase is thought to be robustly known and the flux is almost purely $\nu_e$. The observation of $\nu_e$ burst can rule out some mixing scenarios. Collective effects do not affect this signal, since the absence of $\bar{\nu}_e$ implies that bipolar oscillations do not develop. An exception is, the O-Ne-Mg SN, where MSW resonances may lie deep inside the collective regions. If the resonances are semi-adiabatic, one gets ``MSW-prepared spectral splits'', two for normal hierarchy and one for inverted. Such a signature may be used to determine the neutrino mass hierarchy~\cite{Dasgupta:2008cd}. Additionally, the prompt signal can be used to bound the absolute neutrino mass.

In the later phases of the SN, explosion neutrinos and antineutrinos of all flavors are emitted. However, only $\nu_e$ and $\bar{\nu}_e$ can be detected with significant statistics.  A combination of collective effects and MSW effects produces distinctive signatures in the energy spectra of $\nu_e$ and $\bar{\nu}_e$ , for various mass and mixing scenarios.  A detailed observation has the power to confirm/rule out a number of mixing scenarios~\cite{Choubey:2010up}, as shown in Table \ref{Table1}. The ones that can in principle be distinguished are, i.e. whether $\theta_{13}$ is large or small, and if the hierarchy is normal or inverted. For intermediate values of $\theta_{13}$, the survival probabilities depend on energy as well as the details of SN density profile.

  A spectral split in $\nu_e$ spectrum in the accretion phase would identify the inverted hierarchy~\cite{Duan:2007bt}. Earth matter effects can be identified at a single detector by measuring the wiggles introduced in the spectra due to Earth effects~\cite{Dighe:2003jg}, or by comparison of signals at two detectors~\cite{Dasgupta:2008my}. Either of these observables could lead to identifying different mass and mixing scenarios. One important point is that the sensitivity of SN neutrino observations to the mass hierarchy gets extended to  $\sin^2\theta_{13}\lsim 10^{-5}$ due to the presence of collective effects~\cite{Dasgupta:2008my}.

\subsection{Testing our theories of SN explosion mechanism}
The energy spectrum of the $\nu_e$ burst is modified by collective effects for an O-Ne-Mg SN, and this may be used to identify the SN progenitor~\cite{Dasgupta:2008cd}. Even for iron-core SN, a spectacular signal may be seen if there is an early phase transition in the core, leading to a large $\bar{\nu}_e$ burst, telling us something exciting about stellar astrophysics and hadronic physics~\cite{Dasgupta:2009yj}.

The high-statistics light-curve of the SN can tell us about the different phases of the explosion~\cite{Totani:1997vj}. The presence of stochastic oscillations in the luminosity will be a signal for the predicted SASI modes~\cite{Lund:2010kh}. Detailed studies of the flavor spectrum can point out shock wave effects~\cite{Gava:2009pj} and spectral swaps that distinguish the accretion and the cooling phases~\cite{Choubey:2010up}. Unexpected features such as an abrupt termination of the flux will indicate  black-hole formation~\cite{Beacom:2000qy}.

\section{Conclusions}

The main hurdle in interpreting SN neutrino data, besides the absence of it at present, is the lack of knowledge of initial conditions, i.e. the initial fluxes and densities. This is likely to lead to degeneracies. However, one may expect various aspects of the time and flavor dependent SN signal to be used in synergy. Future developments in SN theory/simulations could be expected to reduce or eliminate some of these degeneracies. Another area of improvement would be a better understanding of the flavor conversion. In particular, the effect of anisotropy and inhomogeneity on collective effects and MSW conversions.

A future galactic SN can be expected to provide a wealth of scientific information for neutrino oscillation physics and SN astrophysics. This is a rare opportunity, occurring once-in-a-lifetime, and we must be ready with suitable detectors and the required theoretical understanding to interpret the data. A significant step towards this, would be a detailed understanding of the rich phenomenology of neutrinos from supernovae.

\section*{Acknowledgements}
I would like to thank my collaborators, especially Amol Dighe, Alessandro Mirizzi and Georg Raffelt for fruitful discussions, and the organizers of the ICHEP 2010 for their hospitality.

\end{document}